\documentstyle[10pt]{article}
\input amssym

\title{Anisotropic fluids in the case of stationary and axisymmetric spaces 
of general relativity}
\author{ T. Papakostas\\ Department of Physics\\ University of Crete,Heraklion,\\710-03 Crete,  
Greece,\\ email: taxiar@physics.uoc.gr\\
Comments:14 pages, submitted to
Inter. J'nal Mod. Phys.}
\begin{document}
\maketitle
\begin{abstract}

We present a stationary axisymmetric solution belonging to
Carter's
familty $[\tilde{A}]$ of spaces and representing an anisotropic fluid 
configuration
\end{abstract}

\begin{abstract}
 We present a stationaryaxisymmetric solution, belonging to
Carter's family $[\tilde{A}]$ of spaces and representing an anisotropic
fluid configuration
\end{abstract}

\section{Introduction}

In the context of general relativity the solutions describing a rotating
fluid body are of great importance. They permit to construct realistic
models of stars and compare the theoretical predictions to actual
observational data concerning rotating stars. Thus the existence of rotating
fluid metrics is the first step for the realization of such a project, the 
second being the matching of the interior solution to an asymptotically flat
vacuum metric at a boundary surface of zero pressure. 

Currently there is no global spacetime model available for a rotating fluid
body in an asymptotically flat vacuum. The problem arises because very few
solutions representing a rotating fluid are known, the most important being
the Wahlquist metric \cite{1},\cite{2}, which has been studied for three decades,
but it has not been succeeded to match this metric to some asymptotically
flat vacuum exterior or to prove the existence of such an exterior metric
rigourously. Recently Bradley et all \cite{3}, have proven the impossibility
of matching the Wahlquist metric to a vacuum exterior in the slow rotation
limit.

I think that such proofs do not tackle the problem for many reasons: (1) the
choice of the extrior vacuum is not unique, (2) a very interesting aspect of the Wahlquist 
metric is that within the range of a parameter the  surface of zero
pressure is prolate along the axis of rotation and this indicate the action
of an external force \cite{4}, then obviously it will be more meaningfull
to match the Wahlquist metric to some other matter metric (that of a null
fluid) and not directly to a vacuum metric.

Also we have to notice that the equation of state for the Wahlquist metric
is $e+3p=const$ and this equation has to be interpreted , otherwise the
physical content of the solution will remain obscure. The possible
directions of research are the following: (1) try to match the Wahlquist 
metric to that of a null fluid and this later to a vacuum exterior, giving
in the same time an interpretation for the equation of state $e+3p=constant$
(2) find new solutions describing rigidly rotating or differentially
rotating \cite{5},\cite{6} perfect fluid and match them to appropriate
vacuum exterior  (3) drop the assumption of perfect  fluid and assume
anisotropy which is suggested by the complexity of the strong interactions
in certain density ranges \cite{7}. The anisotropy of fluids in the context
of General Relativity is already used in the domain of exact solutions 
\cite{8},\cite{9}. Also we have to notice that Florides \cite{10} used the 
Florides-Synge method \cite{11} to show that up to $\kappa^5$ ( where
$\kappa$ is some small parameter) the Kerr metric may be matched to an
interior solution descrbing a rotating body of non perfect fluid (with
anisotropic pressures) and in another paper Florides \cite{12}
used the same method to match the Kerr metric up to $\kappa^5$ to a rigidly 
rotating oblate spheroid with anisotropic pressures.

These results gave us the idea to study the anisotropic fluids in the case
of axisymmetric stationary spaces of General Relativity  and because of the
compexity of Einstein's equations in this case we begin by considering a
more "special case" the Carter's family of metics. In section 2 we present a
brief review of the Newman-Penrose (NP) formalism in the complex vectorial 
representation of Cahen, Debever, Defrise \cite{13},\cite{14},\cite{15} in
which we perform our calculation and we present Carter's family of metrics.
In section 3 we write the energy-momentum tensor of an anisotropic fluid 
and we calculate the components of the Ricci traceless tensor and the Weyl 
tensor, we try then to solve the resulting equations by imposing some kind
of equation of state . In section 4 we study the solution obtained and we
give some perspectives concerning the continuation of our work.

\section{ NP formalism and Carter's family of metrics}

The Carter's family of metrics \cite{16},\cite{17} can be characterised by
the existence of a second rank Killing tensor with two double eigenvalues
$\lambda_1,\lambda_2$ and a two parameter Abelian isometry group $G_2$,
with nonnull surafaces of transitivity with orbits which can be time-like 
or space-like.  We have studied the full family of Carter's metric in the
presence of a perfect fluid \cite{18} and the member $[\tilde{A}]$  of the
family ($\lambda_1,\lambda_2$ are not constant) in \cite{19} where we obtain
a generalization of the Wahlquist solution. In the reference someone can 
find a short resum\'e of the relation between Carter's spaces and the
Hauser-Mahliot spaces and the implications of the existence of a second 
rank Killing tensor on the separability of the Hamilton-Jacobi equation 
for the geodesics. This Killing tensor which characterises Carter' family 
$[\tilde{A}]$ of solutions can be written in a local coordinate system:
\begin{equation}
 K_{ij}=\lambda_1(n_il_j+n_jl_i)+\lambda_2(\bar{m}_im_j+m_j\bar{m}_j)
\end{equation}
where the covariant vectors $l_i$ and $n_i$ are real and null while the 
null vectors $m_i,\bar{m}_i$ are complex conjugate. the functions
$\lambda_1, \lambda_2$ are the two double eigenvalues of the Killing tensor
and they are real. The four vectors form a covariant null tetrad and the
metric can be put in the form 
\begin{equation}
ds^2=2(\theta^1\theta^2-\theta^3\theta^4)
\end{equation}
where
\begin{equation}
\theta^1=n_i dx^i, \theta^2=l_i dx^i, \theta^3=-\bar{m}_idx^i,
\theta^4=-m_idx^i
\end{equation}
A basis for the space of complex self-dual 2-forms is given by 
\begin{equation}
Z^1=\theta^1\wedge \theta^2, Z^2=\theta^1\wedge^2-\theta^3\wedge\theta^4,
Z^3=\theta^4\wedge\theta^2
\end{equation}
The components of the metric in this base are 
\begin{equation}
\gamma^{\alpha\beta}=4(\delta^\alpha_{(1}\delta^\beta_{3)}-
\delta^{\alpha}_2\delta^\beta_2)
\end{equation}
The complex connection 1-forms are defined by 
\begin{equation}
dZ^\alpha=\sigma^\alpha_\beta\wedge Z^\beta
\end{equation}
Greek indices=1,2,3, Latin indices=1,2,3,4 and the vectorial connection
1-form is defined by 
\begin{equation}
\sigma_\alpha=\frac18e_{\alpha\beta\gamma}\gamma^{\gamma\delta}
\sigma^\beta_\delta 
\end{equation}
where $e_{\alpha\beta\gamma}$ is the three dimensional permutation symbol.
The tetrad components $\sigma_\alpha=\sigma_{\alpha a}\theta^a $ are 12
complex valued functions which are exactly the NP spin coefficients:
\begin{equation}
\sigma_{\alpha a}=\left(\begin{array}{cccc}
\kappa & \tau & \sigma & \rho \\
\epsilon & \gamma & \beta & \alpha \\
\pi & \nu & \mu & \lambda 
\end{array}\right)
\end{equation}
The complex curvature  2-forms $\Sigma^\alpha_\beta$ are defined by 
\begin{equation}
d\sigma^\alpha_\beta-\sigma^\alpha_\gamma\wedge\sigma^\gamma_\delta=
\Sigma^\beta_\delta
\end{equation}
and the vectorial curvature 2-form by
\begin{equation}
\Sigma_\alpha=\frac18e_{\alpha\beta\gamma}\gamma^{\gamma\delta}
\Sigma^\beta_\delta
\end{equation}
On expanding $\Sigma_\alpha$ in the basis $[Z^\alpha,\bar{Z}^\alpha]$ one
obtains:
\begin{equation}
\Sigma_{\alpha}=(C_{\alpha\beta}-\frac16R\gamma_{\alpha\beta})Z^\beta +
E_{\alpha\bar{\beta}}Z^{\bar{\beta}}
\end{equation}
where the quantities $C_{\alpha\beta}$ and $E_{\alpha\bar{\beta}}$ are
related to the NP curvature components $\Psi_A$ and $\Phi_{AB}$ as follows:
\begin{equation}
C_{\alpha\beta}=\left(\begin{array}{ccc}
\Psi_0 &\Psi_1 &\Psi_2 \\
\Psi_1 & \Psi_2 &\Psi_3 \\
\Psi_2 & \psi_3 &\Psi_4  
\end{array}\right), \hspace{0.25truecm}
E_{\alpha\bar{\beta}}=\left(\begin{array}{ccc}
\Phi_{00} & \Phi_{01} & \Phi_{02} \\
\Phi_{10} & \Phi_{11} & \Phi{12} \\
\Phi_{20} & \Phi_{21} & \Phi{22}
\end{array}\right)
\end{equation}
The Carter's family $[\bar{A}]$ of metrics can be written as follows
\begin{eqnarray}
ds^2=(\Phi+\Psi)\{\frac{fE^2}{(B-A)^2}(dt+Adz)^2-\frac{H^2}{(B-A)^2}(dt+Bdz)^2
-\frac{f(\Psi_ydy)^2}{4G^2}-\frac{(\Phi_xdx)^2}{4F^2}\}
\end{eqnarray}
where 
\begin{eqnarray}
\lambda=\Phi(x), \lambda_2=\Psi(y),
&\Phi_x=\frac{d\Phi}{dx}& \Psi_y=\frac{d\Psi}{dy}\\
A=A(x), &H=H(x)& F=F(x), \\
B=B(y), &E=E(y)& G=G(y)
\end{eqnarray}
$f=+1$ there is one time-like Killing vector $\frac{\partial}{\partial t}$
and one space like Killing vector $\frac{\partial}{\partial z}$, this is
the axisymmetric case.

\noindent $f=-1$ the Killing vectors are both space-like.

\noindent We consider in this paper only the $f=+1$ case. For all plausible
energy-momentum tensors the traceless Ricci components have to be real, the
only complex components is $\Phi_{01}$ and its imaginary part is equal to 
\begin{eqnarray}
\frac34\frac{i}{(\Phi+\Psi)}\frac{4GF}{\Phi_x\Psi_y}
\{\ln(\frac{\Phi+\Psi}{B-A})\}_{xy}
\end{eqnarray}
where $\{\}_{xy}=\frac{\partial^2}{\partial x \partial y}\{\}$. The vanishing 
of this expression is the necssary and sufficient condition for the
separability of the Scr\"odinger equation:
\begin{eqnarray}
\{\ln(\frac{\Phi+\Psi}{B-A})\}_{xy}=0
\end{eqnarray}
The only solution of (16) that has been used in the literature is the most
obvious one 
\begin{eqnarray}
B(y)=\Psi(y), A(x)=-\Phi(x)
\end{eqnarray}
The general solution of (16) is given by 
\begin{equation}
A(x)=\frac{l_i\Phi+l_2}{l_3\Phi+l_4}, B(y)=\frac{l_1\Psi-l_2}{l_3\Psi-l_4}
\end{equation}
In \cite{19} we claimed that solution (18) could lead in the presence of 
a perfect fluid, to different metrics than that of Wahlquist or the
generalization of Wahlquist obtained there. Unfortunately the calculations
proved that there is no new solution in the case of a perfect fluid energy
momentum tensor. So we present this generalization of the Wahlquist solution
with some minor changes in the constants, which clarify the reduction to the
Wahlquist solution, we present also the vacuum metric of the Carter's family
given by (13) and how we get the Kerr metric from (13) in order to compare
these metrics with the obtained for the anisotropic fluids.

We generalize the Wahlquist solution as follows
\begin{eqnarray*}
ds^2=\frac{1}{\zeta^2+\xi^2}\{E^2(\zeta)[dt-(l\xi^2-\frac{k_2-l}{2q^2})dz]^2
&-H^2(\xi)[dt+(l\zeta^2-\frac{k_2-l}{2q^2})dz]^2\}&-  \\ 
-(\zeta^2+\xi^2)
[\frac{d\zeta^2}{E^2(\zeta)(1-q^2\zeta^2)}+\frac{d\xi^2}{H^2(\xi)(1+q^2\xi^2)]}
\end{eqnarray*}

\begin{equation}
E^2(\zeta)=-\frac{a^2}{q^2}\zeta(1-q^2\zeta^2)^\frac12
\sin ^{-1}(q\zeta)+(B+2\Gamma q^2) 
\zeta^2+p_1\zeta(1-q^2\zeta^2)^\frac12-\Gamma,
\end{equation}
\begin{equation}
H^2(\xi)=\frac{a^2}{q^2}\xi(1-q^2\xi^2)^\frac12
\sinh ^{-1}(q\xi)-(B+2\Gamma q^2)\xi^2+p_2\xi(1+q^2\xi^2)^\frac12-\Gamma
\end{equation}
and 
\begin{equation}
B=\frac{1}{\beta}(a^2+\frac{\gamma_1+1}{\gamma_2r_0^2})
\end{equation}

\begin{equation}
\Gamma=\frac{1}{\beta^2(\beta+1)\gamma_2r_0^2}[a^2(1-\beta)\gamma r_0^2 +
2(\gamma_1-\beta)]
\end{equation}

\begin{equation}
k_2=\gamma_2 r_0^2, l=\gamma_2\beta r_0^2
\end{equation}

\begin{equation}
\beta=(1-4q^2)^\frac12
\end{equation}
and $\gamma_1=\gamma_1(a,q), \gamma_2=\gamma_2(a,q)$ 
\begin{equation}
\lim_{a,q\rightarrow 0}\gamma_1=0, \lim_{a,q \rightarrow 0}\gamma_2=1
\end{equation}
The constants of this metric are $a,q,r_0,p_1,p_2,\gamma_1,\gamma_2$.
Obviously $\gamma_1,\gamma_2$ are not independent constants, they depend on
$a,q$ which are the fluid constants: if $a=q=0$ the fluid disappears, 
$r_0$ is the radius of the system of oblate spheroidal coordinates
and if $p_1=p_2=0$ there is no ring singularity.

We get the Wahlquist solution setting 

\begin{equation}
\gamma_2=\frac{2(1-\beta)}{\beta(\beta+3)-a^2r_0^2(1-\beta)},
\gamma_1=\frac{(1-\beta)[\beta^2(\beta-2)-a^2r_0^2(1+2\beta)]}{\beta(\beta+3)-
a^2r_0^2(1-\beta)}
\end{equation}
\begin{equation}
\xi_A^2=\frac{2}{\beta(\beta+1)},\hspace{0.1truecm}
 \delta=\pm\beta\gamma_2 r_0^2, \hspace{0.1truecm}
z=\tilde{z}r_0,\hspace{0.1truecm} a^2 r_0^2=\frac{1}{k^2}
\end{equation}
$\xi_A,\delta $ are the constants used by Wahlquist, they are not
independent, but Wahlquist did not show explicitly their dependence in his 
paper.

The relation of the Wahlquist coordinates $\zeta,\xi$ with $x,y$ of the
metric (13) is
\begin{equation}
x^2=l\xi^2-\frac{k_2-l}{2q^2}, \hspace{0.25truecm}
y^2=l\zeta^2+\frac{k_2-l}{2q^2}
\end{equation}
The vacuum solution for the metric (13) is given by:
\begin{equation}
B(y)=\Psi(y)=y^2, G^2=y^2E^2(y), 
A(x)=-\Phi(x)=-x^2, F^2=x^2H^2(x)
\end{equation}
\begin{equation}
E^2(y)=\frac12 ay^2+by+c, H^2(x)=-\frac12ax^2+dx+c
\end{equation}

\noindent The Kerr metric can be obtained from (28) if we set:
\begin{equation}
a=2,d=0,b=-2m,c=\alpha^2\\
y=r, x=\alpha\cos \theta
\end{equation}
$r,\theta$ are the Boyer, Lindquist coordinates then we have that 
\begin{equation}
E^2=r^2-2mr+\alpha^2,
H^2=\alpha^2\sin ^2\theta
\end{equation}
\begin{equation}
x^2+y^2=r^2+\alpha^2\cos^2\theta
\end{equation}
and the definition of new coordinates $\hat{t},\phi$:
\begin{equation}
dz=ad\phi, \hspace{0.4truecm} dt=d\hat{t}+ d\phi
\end{equation}
brings the metric to its final form.

\section{ Carter's metric $[\tilde{A}]$ in the presence of an anisotropic
fluid}

We assume that the energy momentum tensor is locally anisotropic and in the 
tangent space can be put in the form:
\begin{equation}
T^i_j=\left(\begin{array}{cccc}
         \rho & 0 & 0 & 0 \\
         0    & -p_y & 0 & 0 \\
         0    & 0  & -p_z & 0 \\
         0    & 0  &  0 &  -p_x 
        \end{array}\right)
\end{equation}
This is most general case for a second rank tensor, it appears to have four 
distinct eigenvalues (one positive and three negative), if we use the tetrad 
defined by (3), we can write the energy momentum tensor as follows:

\begin{eqnarray}
T_{ij}=\frac12(e+p_y)(n_i n_j+l_i l_j)+\frac12(p_z-p_x)(\bar{m}_i\bar{m}_j 
+m_im_j)+\nonumber \\
+\frac14(e-p_y+p_z+p_x)(n_il_j+l_in_j+\bar{m}_im_j+m_i\bar{m}_j)+\nonumber\\
\frac14(e-p_y-p_z-p_x)(n_il_j+l_in_j-\bar{m}_im_j-m_i\bar{m}_j)
\end{eqnarray}
The Einstein equations 
\begin{equation}
R_{ij}-\frac12Rg_{ij}=T_{ij}
\end{equation}
give the following expressions for the components of the traceless Ricci
tensor in the NP notation:
\begin{eqnarray}
\Phi_{01}=\Phi_{12}=0\\
\Phi_{00}=\frac14(e+p_y),&& \Phi_{02}=\frac14(p_z-p_x)\\
2\Phi_{11}=\frac14(e-p_y+p_z+p_x)&,& 6\Lambda=\frac14(e-p_y-p_z-p_x)
\end{eqnarray}
The Carter metric $[\tilde{A}]$ can be written in the following way:
\begin{eqnarray}
ds^2=\frac{1}{x^2+y^2}\{ E^2(y)[dt-x^2dz]^2-H^2(x)[dt+y^2dz]^2\}-\nonumber\\
-(x^2+y^2)[\frac{y^2dy^2}{G^2}+\frac{x^2dx^2}{F^2}]\}
\end{eqnarray}
The components of the traceless Ricci tensor and the Weyl tensor in the NP
notation for the metric (36) are:
\begin{eqnarray}
\Phi_{01}=\Phi_{21}&=&\frac{HE}{4(x^2+y^2)^3}[\frac{(T^2)_y}{2y}-
\frac{(\Pi^2)_x}{2x}],\\
\Phi_{00}=\Phi_{22}&=&\frac{E^2}{2(x^2+y^2)^3}[\frac{(T^2+\Pi^2)}{x^2+y^2}-
\frac{(T^2)_y}{2y}]\\
\Phi_{02}=\Phi_{20}&=&\frac{H^2}{2(x^2+y^2)^3}[\frac{(T^2+\Pi^2)}{x^2+y^2}-
\frac{(\Pi^2)_x}{2x}]\\
\Psi_1=\Psi_3&=&\frac{HE}{2(x^2+y^2)^3}[-\frac{2(T^2+\Pi^2)}{x^2+y^2}+
\frac{(T^2)_y}{2y}+\frac{(\Pi^2)_x}{2x}]\\
\Psi_0=\Psi_4&=&0
\end{eqnarray}

\noindent $\Phi_{11},6\Lambda,\Psi_2$ are lengthy and will not be given here,
we have used the notations:
\begin{eqnarray}
T(y)=\frac{G(y)}{E(y)}, \pi(x)=\frac{F(x)}{H(x)}, \hspace{0.15truecm}
(T^2)_y=\frac{d T^2}{dy} \dots
\end{eqnarray}
If we impose conditions (35) to the expressions (37)-(41) we get that:
\begin{eqnarray}
\frac{(T^2)_y}{2y}-\frac{(\Pi^2)_x}{2x}=0
\end{eqnarray}
The solution of this equation is
\begin{eqnarray}
T^2(y)=\frac{G^2(y)}{E^2(y)}=k_2y^2+k_0, \Pi^2(x)=\frac{F^2(x)}{H^2(x)}=
k_2x^2+l_0
\end{eqnarray}
Substitution of (44) in (38)-(40) implies that:
\begin{eqnarray}
\Phi_{00}=\Phi_{22}&=&\frac{E^2(y)(k_0+l_0)}{2(x^2+y^2)}\\
\Phi_{02}=\Phi_{20}&=&\frac{H^2(x)(k_0+l_0)}{2(x^2+y^2)}\\
\Psi_1=\Psi_3&=&-\frac{H(x)E(y)(k_0+l_0)}{2(x^2+y^2)}
\end{eqnarray}

\noindent $2\Phi_{11},6\Lambda$ are given in appendix. Expression (47) permit
the following classification
\begin{eqnarray}
\Psi_1=\Psi_3=0 &\mbox{or}&  k_0+l_0=0, \Psi_2\neq 0\\
\mbox{and}\\
\Psi_1=\Psi_3\neq 0 , &k_0+l_0\neq 0&\Psi_2\neq 0
\end{eqnarray}

Relations (48) imply that the metric is of type D in the Petrov classification
and it is determined by 928), this metric is the vacuum case of Carter's
family $[\tilde{A}]$.  Relations (49) imply that the metric is of type I in
the Petrov classification provided that $9\Psi^2_2\neq 16\Psi_1^2$, this
metric has been obtained by imposing the anisotropic fluid energy momentum 
tensor (33). It is obvious that in our approach we have still two unknown
functions $E^2(y),H^2(x)$ and no other equation available to define them. We
have to impose a supplementary condition and this condition could be an
equation of state between the pressure and the rset energy $e$, but we have
to pay attention to the fact that the quantities $p_x,p_y,p_z$ contain 
contribution from fluid pressure as well as other stresses and that
generally these quantities depend on additional variables (such as entropy
magnetic fields etc). The supplementary condition that we impose is
justified only by the fact that we can write down an equation which can be
solved and define in this way $E^2(x),H^2(x)$:
\begin{eqnarray}
e+p_z=2(p_x-p_y)
\end{eqnarray}
Using (50) , (35) and the expressions (45),(46) and those of the appendix
of the appendix we can write the following equation:
\begin{eqnarray}
(y^2+x^2)(k_2y^2+k_0)\frac{(E^2)_{yy}}{y^2}-
(4k_2y^4+5k_0y^2+k_0x^2)\frac{(E^2)_y}{y^3}+4k_2E^2-   \nonumber\\
-(y^2+x^2)(k_2x^2+l_0)\frac{(H^2)_{xx}}{x^2}-
(4k_2x^4+5k_0x^2+l_0y^2)\frac{(H^2)_x}{x^3}+4k_2H^2=0
\end{eqnarray}

\noindent If we differentiate twice (51) with respect to x and twice with respect to y
we get the separation of $x$ and $y$
\begin{eqnarray}
(k_2y^2+k_0)\frac{(E^2)_{yy}}{y^2}-k_0\frac{(E^2)_y}{y^3}&=&
-\frac{d}{2n}y^2+a_1y+a_0,\\
(k_2x^2+l_0)\frac{(H^2)_{xx}}{x^2}-l_0\frac{(H^2)_x}{x^3}&=&
-\frac{d}{2n}x^2+a_1x+a_0
\end{eqnarray}
the constant $d$ is the constant of separation and the constants
$a_1,a_0,b_1,b_0,n$ are constants of integration. the integration by parts
of (52),(53) result two linear differential equations of first order:

\begin{eqnarray}
(k_2y^2+k_0)(E^2)_y-k_2yE^2=-\frac{d}{8n}y^5+\frac13a_1y^4+\frac12 a_0y^3+
r_1y
\end{eqnarray}
\begin{eqnarray}
(k_2x^2+l_0)(H^2)_x-k_2xH^2=-\frac{d}{8n}x^5+\frac13b_1x^4+\frac12b_0x^3+
r_2x
\end{eqnarray}
These equations can be easily integrated and the expressions for $E^2,H^2$ 
are substituted in (51) which has to be satisfied identically. Finally we
get:
\begin{eqnarray}
E^2=q_1\sqrt{k_2y^2+k_0}-\frac{d}{24k_2\eta}y^4+\frac{1}{2k_2}
(\frac{k_0d}{3k_2\eta}+a_0)y^2-\frac{r_1}{k_2}+\frac{k_0^2d}{3k_2^3\eta}+
\frac{a_0k_0}{k_2^2}
\end{eqnarray}
\begin{eqnarray}
H^2=q_2\sqrt{k_2x^2+l_0}-\frac{d}{24k_2\eta}x^4+\frac{1}{2k_2}
(\frac{l_0d}{3k_2\eta}-a_0)x^2-\frac{r_1}{k_2}+\frac{l_0^2d}{3k_2^3\eta}-
\frac{a_0l_0}{k_2^2}
\end{eqnarray}
where $q_1,q_2,d,k_2,\eta,r_1,r_2,a_0,k_0,l_0$ are arbitrary constants.
\section{ Properties of the solution}

Our approach of anisotropic fluids is based on the form of the
energy-momentum tensor (38). This form has the following disadvantages: 
there is no natural way to define  the hydrostatic pressure and
consequently it is not possible to obtain the zero pressure surface which 
is necessary ary to construct a realistic star model, also we have no
four-velocity for a comoving with the fluid observer so we are not able to
characterize the motion of the fluid.

On the other hand this solution (metric (43) with (51) and (64),(65) ) it
is the first exact solution which can describe an anisotropic fluid. 
Wahlquist in \cite{2} mentioned the possiblity for such solutions but he has
not succeeded to solve the differential equations which are satisfied by the 
two remaining unknown functions. Finally this solution reduces to the
vacuum family $[\tilde{A}]$ of Carter's spaces if we impose the condition:
\begin{eqnarray}
k_0+l_0=0
\end{eqnarray}
Some evidence that our solution describes a realistic anisotropic fluid
comes from the fact that it is compatible with the existence of a rotation
axis which satisfies the condition of elementary flatness. Also our solution
can in principle satify the strong energy conditions:

The position of the rotation axis is given by the vanishing of the axial
Killing vector:
\begin{equation}
x=0 \hspace{0.2truecm} \mbox{and} \hspace{0.2truecm} H^2(x=0)=0
\end{equation}
these relations imply that 
\begin{eqnarray}
q_2\sqrt{l_0}-\frac{r_1}{k_2}+\frac{l_0^2d}{3k_2^3}-\frac{a_0l_0}{k_2^2}=0
\end{eqnarray}
and that 
\begin{eqnarray}
l_0>0
\end{eqnarray}
The elemntary flatness of the rotation axis is quarranteed by the regularity
condition:
\begin{eqnarray}
\frac{X_{,i}X^{,i}}{4X}\rightarrow 1, \hspace{0.5truecm} \mbox{on the axis}
\end{eqnarray}
where 
\begin{eqnarray}
X=U_iU^i , \hspace{0.5truecm} U=\frac{\partial}{\partial z}
\end{eqnarray}
It is remarkable that relation (60) ensures (62) also!!

If finally we consider expressions for the eigenvalues of the energy-
momentum tensor (given in appendix) we can prove the following statement.
If we suppose that:
\begin{eqnarray}
e+p_z=g_1^2, g_\in \bf{R}\\ e+p_y=g_2^2, g_2 \in \bf{R}
\end{eqnarray}
then we can show that:
\begin{eqnarray}
e+p_x=g_2^2+\frac12g_1^2\\ 
e+p_x+p_y+p_z=g_1^2-\frac14d
\end{eqnarray}
The positivity of the expressions $e+p_z,e+p_y,e+p_x$ and $e+p_x+p_y+p_z$ is
nothing else but the strong energy conditions!! Then we have proved that the
satisfaction of the two of the strong energy conditions impies the validity
of the remaining two conditions (for $d<0$ this true also for all possible
values of $g_1$). 

Obviously we have solved a part of the problem of finding a global space time
model for a rotating fluid. we have to define now in a consistent way a
surface of zero pressure and then match our solution of anisotropic fluid
 to the vacuum spaces of Carter's family $[\tilde{A}]$. The vacuum spaces 
$[\tilde{A}]$ of Carter have not been srtudied until now despite the fact 
that they are an important generalization of Kerr metric and this necessity
is stipulated by Carter in \cite{16}. We think that the existence of four
arbitrary constants in Carter's spaces $[\tilde{A}]$ makes them more
appropriate for the matching with an interior metric. Finally the
determination of the zero pressure surface will permit us to answer if there
is a singularity (the $x=y=0$ singularity) inside the physical region of the
fluid.

\newpage

\section{APPENDIX}

\begin{eqnarray}
2\Phi_{11}=\frac{(k_2y^2+k_0)
\frac{d^2E^2}{dy^2}}{4y^2(x^2+y^2)}
-\frac{(4k_2y^4+5k_0y^2+k_0x^2)
\frac{dE^2}{dy}}{4y^3(x^2+y^2)^2}+
\frac{(2k_2(x^2+y^2)+3(l_0+k_0))E^2(y)}{2(x^2+y^2)^3}+\nonumber\\
-\frac{(k_2x^2+l_0)\frac{d^2H^2}{dx^2}}{4x^2(x^2+y^2)}+
\frac{(l_0y^2+4k_2x^4+5l_0x^2)\frac{dH^2}{dx}}{4x^3(x^2+y^2)^2}-
\frac{(2k_2(x^2+y^2)+3(l_0+k_0)) H^2(x)}{2(x^2+y^2)^3}
\end{eqnarray}

\begin{eqnarray}
-6\Lambda=\frac{(k_2y^2+k_0)
\frac{d^2E^2}{dy^2}}{4y^2(x^2+y^2)}-
\frac{k_0\frac{dE}{dy}}{4y^3(x^2+y^2)}
+\frac{(l_0+k_0)E^2(y)}{2(x^2+y^2)^3}+\nonumber \\
\frac{(k_2x^2+l_0)\frac{d^2H^2}{dx^2}}{4x^2(x^2+y^2)}
-\frac{l_0\frac{dH}{dx}}{4x^3(x^2+y^2)}
-\frac{(k_0+l_0)H^2(x)}{2x^3(x^2+y^2)^3}
\end{eqnarray}

\end {document}